\documentclass[a4paper]{jpconf}

\usepackage{graphicx}
\usepackage{citesort}
\usepackage{amsmath}
\usepackage{wrapfig}
\usepackage{url}

\begin{document}
\title{%
A development of an accelerator board dedicated for 
multi-precision arithmetic operations and its application 
to Feynman loop integrals II %
}

\author{
H~Daisaka$^1$,
N~Nakasato$^2$,
T~Ishikawa$^3$,
F~Yuasa$^3$,
K~Nitadori$^4$
}

\address{$^1$ Hitotsubashi University, 2--1, Naka, Kunitachi, Tokyo, 186--0801, Japan}
\address{$^2$ University of Aizu, Aizu-wakamatsu, Fukushima,  965--8580, Japan}
\address{$^3$ High Energy Accelerator Research Organization (KEK), 1--1, Oho, Tsukuba, Ibaraki, 305--0801, Japan}
\address{$^4$ RIKEN Advanced Institute for Computational Science, 7--1--26, Minatojima-minami-machi, Chuo-ku, Kobe, Hyogo, 650--0047, Japan}

\ead{daisaka@phys.science.hit-u.ac.jp}

\begin{abstract}
Evaluation of a wide variety of Feynman diagrams with multi-loop integrals 
and physical parameters and its comparison with high energy experiments 
are expected to investigate new physics beyond the Standard Model.
We have been developing a direct computation method of 
multi-loop integrals of Feynman diagrams. 
One of features of our method is 
that we adopt the double exponential rule for numerical integrations
which enables us to evaluate loop integrals with boundary singularities.
Another feature is that 
in order to accelerate the numerical integrations
with multi-precision calculations, 
we develop an accelerator system with Field Programmable Gate Array boards 
on which processing elements with dedicated logic 
for quadruple/hexuple/octuple precision arithmetic operations 
are implemented.
In addition, we also develop a programming interface designed 
for easy use of the system. 
The development is continued for practical use of the system.
We present the current development of our system, and
the numerical results of higher-loop diagrams performed using our system.
\end{abstract}
\section{Introduction}

The discovery of the Higgs boson in 2012 opened the door to a new era of 
elementary particle physics. While the Standard model has been successful, 
we have good enough reasons to expect new phenomena beyond it.
In these investigations, precise theoretical predictions are crucial.

Recently,
methods of evaluating Feynman integrals have expanded greatly 
in both analytic and numerical approaches. 
When we go into the higher order calculation, we have to include over 
thousands of diagrams and thus an automatic 
computation system is strongly required.
Enormous efforts for developing such systems have been dedicated by many 
authors. 
For 1-loop integrals, several automated systems have shown
successful results~\cite{HAHN1999153,Wang_2004,Belanger_2006,Cullen_2011,Hameren_2011,Carrazza_2016,Donner_2017}.
For multi-loop integrals, a lot of methods have been developed and the current status is reviewed
in~\cite{Borowka_2017}.
However, for evaluation of a wide variety of integrals including propagators 
with arbitrary mass scale and external momentum, there is still room for 
improvement.
%
%
%
%
%

We have been developing a fully numerical method for the evaluation of 
Feynman integrals which called 
{\it Direct Computation Method} (DCM)~\cite{Yuasa_2012}.
It is based on the numerical multi-dimensional integration and the 
extrapolation method. 
We can choose any kind of numerical integration method in DCM 
if it gives integration results in a good enough precision. 
Here we use 
the {\it Double Exponential Formula} (DE)~\cite{TM74} which is one of 
the optimal methods in the case that there is an endpoint singularity.
For the extrapolation, we use both linear and nonlinear extrapolation methods.
So far, DCM have succeeded evaluating Feynman integrals with 2-, 3-, 
and 4-loops with 2, 3, and 4 legs~\cite{Doncker_2018}.
%
%

%

%
However, DCM still has some difficulties for evaluating multi-loop diagrams. 
One is that the amount of calculation is very large, 
since a multi-loop diagram becomes an integration of multi-dimension. 
For example, a 3-loop self-energy diagram needs 7-th dimensional integration. 
Second is that an evaluation of multi-loop diagrams often requires
calculation with high accuracy, higher than double precision,
especially in the case that there is a strong singularity.
Such a high accuracy calculation of Feynman loop integrals
can be carried out by using high precision 
softwares (e.g., ~\cite{GMP,QDLIB,MPFR}).
However, computation time will be much longer than
the case using only double-precision.
This is because in such softwares a high accuracy is realized by
a combination of amount of arithmetic operations in ordinary CPUs,
in other words, a large increase in operation count in the high precision case.

As one of the solutions to overcome these problems,
we have been developing a dedicated accelerator system for multi-precision 
arithmetic operation.  
Based on the idea of GRAPE project(e.g.,~\cite{Sugimoto_1990}\cite{Makino_2006}), 
we have designed GRAPE-MPX architecture and the processor 
in which there are a number of processing elements (PE) 
with dedicated logic units for quadruple~(hereafter MP4), 
hexuple~(MP6), octuple-precision~(MP8) arithmetic.
So far, this processor has been implemented on a structured ASIC~\cite{Daisaka_2011} 
and FPGAs~\cite{Nakasato_2012}. 
The latest system is called 
GRAPE9-MPX~\cite{Motoki_2014}\cite{Daisaka_2015}. 

%
In order to advance 
the evaluation of Feynman integrals,  
we continue to develop GRAPE9-MPX to have much computational power.
We also continue to develop our programming interface, Goose and LSUMP, 
to fit to the current system, which supports the use of MPI and, furthermore, 
the use of GPU without changing our application program. 

This paper organized as follows. 
The current status of GRAPE9-MPX system is explained 
in Section~2, 
and Section~3 presents the results of a 3-loop Feynman integral 
with two legs for the case propagators 
have two different mass scales as an application.
The last section is dedicated for summary and discussion.

\section{Overview of GRAPE9-MPX and programming interface}

Figure~\ref{g9mpx} shows a schematic diagram and a picture of 
GRAPE9-MPX cluster system in KEK.
This system is developed from the previous system with 
a single host and multiple FPGA boards 
\cite{Motoki_2014}\cite{Daisaka_2015} 
in order to enhance the computational power.
The GRAPE9-MPX  cluster system has 64 FPGA boards 
which are installed on 8 host computers. 
Each board is connected to a host via PCIe interface.
For FPGA boards, we used Altera Arria V board from   
Intel Co.  \cite{AlteraArriaVboard}.
With the current implementation of our processor,
the peak performance of the system 
achieves 403 GFlops for MP4, 185 GFlops for MP6, and 96 GFlops for MP8.
{
\setlength\abovecaptionskip{-5pt}
\begin{figure}[h]
\begin{center}
\includegraphics[scale=0.25]{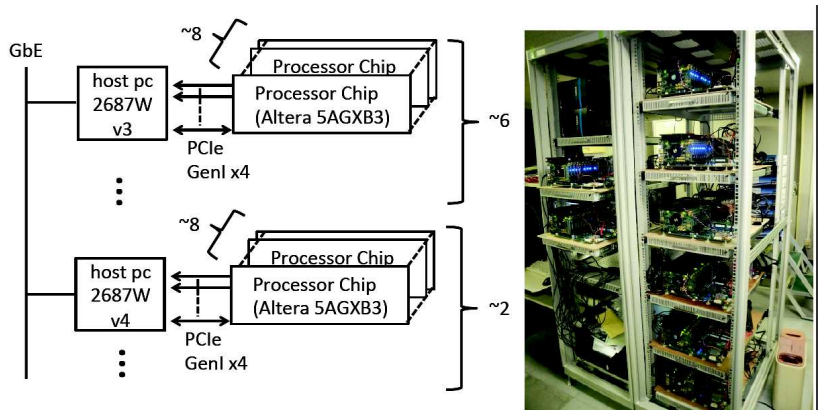}
\caption{Schematic diagram and picture of GRAPE9-MPX cluster system. 
Each host has 8 FPGA boards and is connected with GbE. 
For host PCs, we use Intel Xeon E5-2687W v3(6 hosts) 
and E5-2687W v4(2 hosts) for CPU, 128GB memory, 
and X10 DRX MB from SuperMicro which has 11 PCIe slots. 
\vspace{-0.5cm}
}
\label{g9mpx}
\end{center}
\end{figure}
}
%


Our dedicated processor is designed by using 
VHDL (VHSIC Hardware Description Language)
so that it is possible to implement our processor on 
another FPGA board, and also it is easy to make improvements.
Our processor consists of two parts,
MP processor and Control processor (CP).
MP processor consists of PEs and broadcast memory units 
which form a SIMD processor. 
Each PE has a multiply unit, an adder unit, and 
register units for quadruple/hexuple/octuple-precision arithmetic.
These arithmetic units perform in every clock cycle,
but have 4 clock latency. 
Also, PE has an inverse square root (rsq) unit with a limited accuracy 
(19 bits for exponent and 32 bits for mantissa).
This unit is used to 
create an initial guess for division arithmetic
which is performed by Newton-Raphson method.

The role of CP is to control MP processor,  
by sending data and instruction necessary for calculation 
from a host computer to MP processor,
and receiving calculation results from MP processor and
sending them to the host.
It consists mainly of IO units
(including PCIe and DRAM controllers), 
memory components for data and instruction, 
and their control unit.
In the current implementation,
the data memory has 32k words stored on onchip memory, 
whereas 
the instruction memory has 4k words on onchip memory,
and 16M words on DRAM equipped on the FPGA board.

%
Table~\ref{format_spec} shows a numerical representation 
for MP4, MP6, and MP8 used in PE, 
and a specification of our dedicated processor of the current implementation.
Note that we used 19 bits for exponent in order to follow 
IEEE binary256 format for MP8, and  
the same exponent is also used for MP4 and MP6 for simplicity.
Our numerical representation has the total bit width 4 bits wider 
than a standard format used in a host computer.
Therefore,
we cut 4 bits from mantissa or exponent
when the data is sent back to CP and a host computer.
The environmental variable decides from which we cut the bits.
%

%

\begin{table}[h]
\begin{tabular}{cccc} 
{}  & \shortstack{MP4\\ (quadruple)} & \shortstack{MP6\\ (hexuple)} & \shortstack{MP8\\ (octuple)} \\ \hline
sign      &   1 &   1 &   1 \\
exponent  &  19 &  19 &  19 \\
mantissa  & 112 & 176 & 240 \\ \hline  
total     & 132 & 196 & 260 \\ \hline
standard  & 128 & 192 & 256 \\
\end{tabular}
\begin{tabular}{llll} 
{}           & MP4  & MP6 & MP8 \\ \hline
Number of PE &  36  & 19  & 11  \\
Clock(MHz)   &  88  & 78  & 68  \\
Peak(Gflops) &  6.3 & 2.9 & 1.5 \\ \hline  
logic utilization  (\%) & 96  & 89 & 83 \\ 
mem   utilization  (\%) & 52  & 55 & 62 \\ 
DSP   utilization  (\%) & 64  & 69 & 62 \\ \hline
\end{tabular}
\caption{Numerical representation (length of bit) used in PE (left),
and a specification and resource utilization 
of our processor currently implemented 
in the FPGA board (right).
For logic synthesis, we use Quartus 13.1 by Altera.}
\label{format_spec}
\end{table}

As already stated in \cite{Motoki_2014},
one of applications suited for our system 
is an interaction type calculation as
$f_i = \sum_{j=1}^{n_j}f(X_i,Y_j)$,
where $X_i$ is the $i$-th element of $X$,
$Y_j$ is the $j$-th element of $Y$,
and $n_j$ is the number of elements of $Y$.
At calculation,
data $X_i$ is set on registers in $i$-th PE, 
and then data $Y_j$ is sent from CP to all PEs.
According to instructions sent from CP, 
$i$-th PE performs an evaluation of the function 
$f(X_i, Y_j)$ and a summation of the result. 
This process continues until data $X_i$ and $Y_i$ run out.
In a program, this can be written as double loops.  
Note that a multi-dimensional integration can be expressed 
in double loops by using loop fusion technique
that merges multiple loops into a single loop.
Therefore, it is possible to accelerate a calculation of 
Feynman loop integrals by using GRAPE9-MPX.
%

{
\setlength\abovecaptionskip{-3pt}
\begin{figure}[h]
\begin{center}
\includegraphics[scale=0.25]{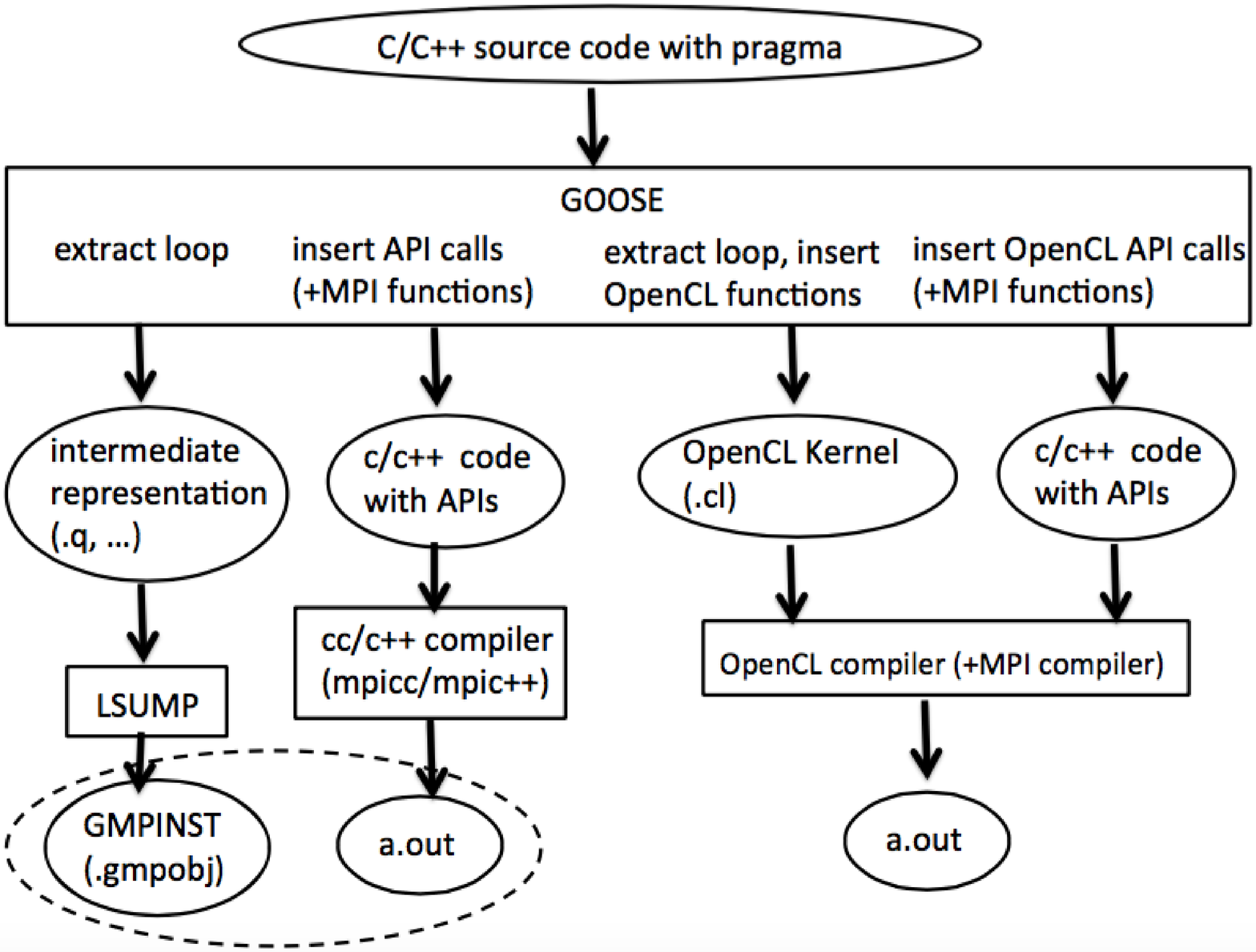}
\includegraphics[width=8.0cm]{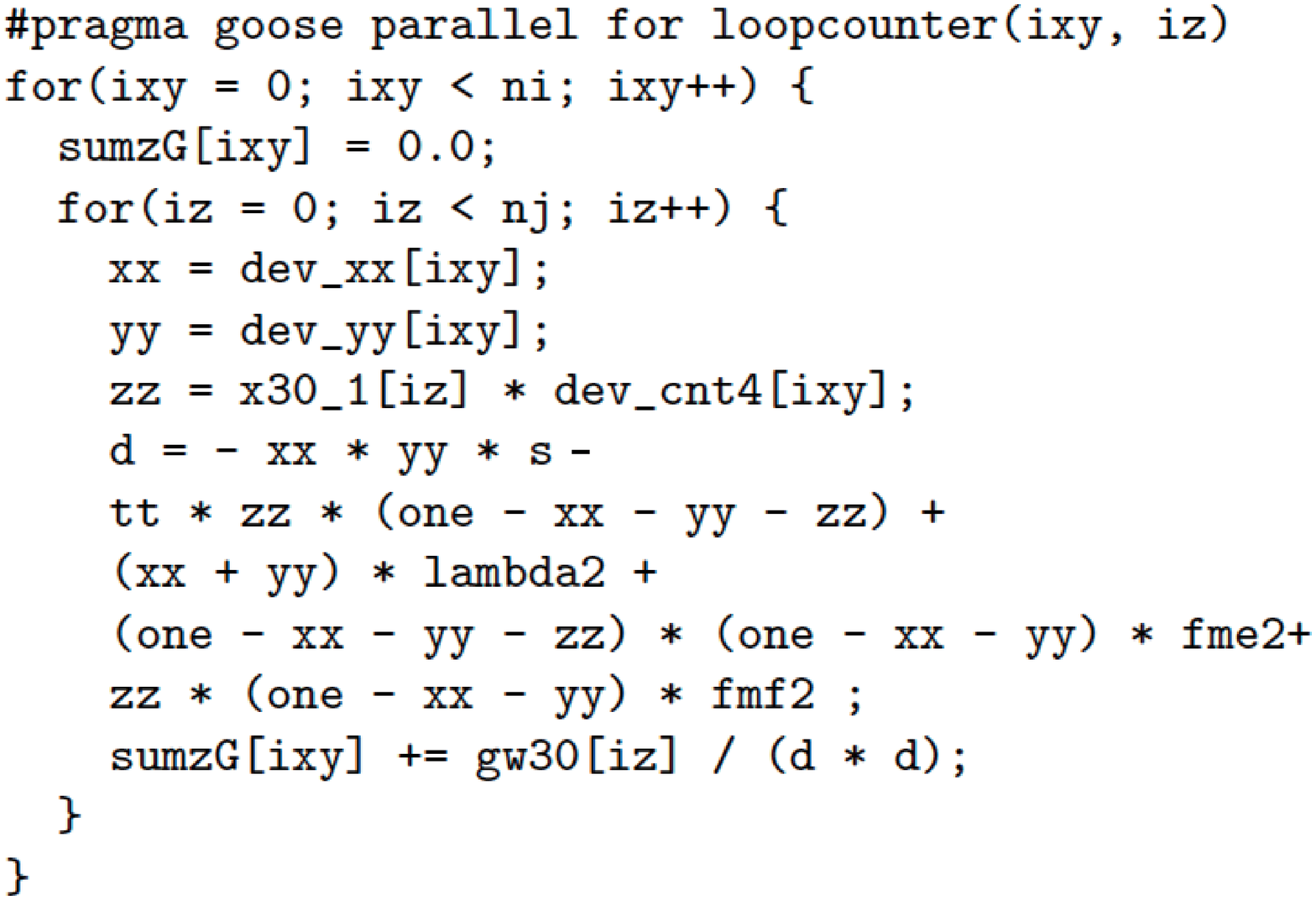}
\caption{Schematic picture of flow of Goose and LSUMP and a 
sample code with the directive for a case of 1-loop box scalar integral. 
\vspace{-0.5cm}
}
\end{center}
\label{goose2017}
\end{figure}
}
%
%

Along with the development of the dedicated accelerator, 
we also have been developing our own compiler system, Goose and LSUMP, 
in order to  use our system easily. 
Figure~\ref{goose2017} shows the flow of compiling a program 
by Goose and LSUMP.
Goose is a directive base compiler like OpenMP. 
Only procedure we have to take is to put a directive pragma 
(\verb|#pragma goose parallel|)
in an original C/C++ source code 
just before a double loop to be accelerated.
LSUMP is our Domain Specific Language (DSL) compiler 
specially developed for GRAPE9-MPX. 
It generates a kernel executed in GRAPE9-MPX from the double loop 
extracted by Goose. 
For more detail, see Nakasato~\cite{Nakasato_2009}. 

%
%
In order to use our cluster with multiple hosts and FPGA boards,
we need to use a parallel computing interface such as MPI. 
We extended Goose to generate MPI API calls as well.
In the current implementation, in the case that $n$ hosts 
are used in calculation in parallel, 
the outer loop ($i$-loop) is divided into $n$ sub loops, 
and each sub loop is assigned to each host. 
Also, we extended Goose to generate OpenCL API calls 
and kernels which enable us to accelerate our program by using GPGPU.
For GPGPU, we use the Double-Double(DD) format~\cite{Hida_2001}. 
Thus, by using Goose,
we can accelerate our application program written in a simple C/C++ source 
code not only on GRAPE9-MPX but also on GPGPU with MPI. 

%
%
%

%
%
%
\section{Application to 3-loop Feynman integral}


%
Here 
we present the numerical results of a 3-loop Feynman integral with 2 legs
by DCM using GRAPE9-MPX system. 
Figure~\ref{fig:loop} shows a diagram of 3-loop self energy and 
the corresponding integral we evaluated.
The number of the propagators is 8 as shown in Fig.~\ref{fig:loop} 
and the number of dimensions of the integration becomes 7 
due to the $\delta$-function. 
The masses of the propagator are given as 
$m_1=m_2=m_5=m_6=m_7=m_8=1/2$ and $m_3=m_4=1$,
in order to compare results 
of Ghinculov~\cite{Ghinculov_1996}.
We proceeded the evaluation of the diagram by changing the kinematic variable $p$
not only in unphysical ($0 \le p^2 < 1$) but also in physical ($1 < p^2 <4$ and $4 <  p^2 \le 8 $) region.
For the physical region, we extrapolated a resultant value 
from results of the integral with different values of $\rho$. 
For the extrapolation,  
we used Wynn's $\epsilon$-algorithm~\cite{Wynn_1956}.
%
Table~\ref{tab:condition} lists a condition of calculation related to 
the DE formula and the extrapolation.
In order to perform an integration that may be affected by singularity,  
we used a larger number of grids and small mesh sizes,
and the extrapolation for calculations in physical region.

{
\begin{figure}[h]
\begin{center}
\includegraphics[scale=0.3]{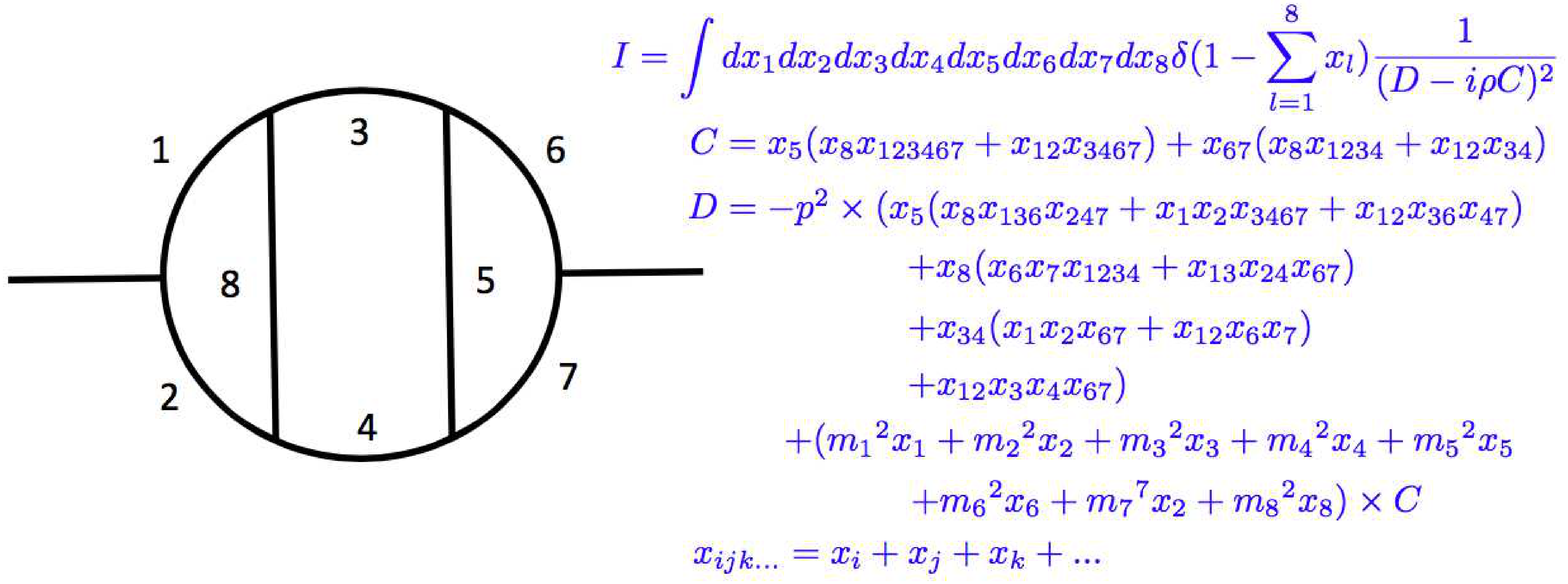}
\caption{3-loop self energy diagram and the corresponding integral
in a massive case we evaluated.
The values $p$ and $x_i$ denote a momentum of an external line and 
Feynman parameters, respectively.
The term $i\rho$ prevents the denominator from vanishing in the integral domain.
}
\label{fig:loop}
\end{center}
\end{figure}
}

%

\begin{table}[h] 
\begin{center}
\begin{tabular}{lll}  \hline
{}        & unphysical region($I_{re}$) & physical region ($I_{re}, I_{im}$) \\ \hline
$N$       &        25         & 64               \\
$h$       &        0.3        & 0.0875, 0.11718  \\
$\rho$    &   0 (fixed)       & $1.15^{\beta}$, $\beta = -51, -52, .., -74$  \\
number of FPGA boards  & 1       & 64 \\
precision              & MP4     & MP4 \\
elapsed time per point & 300 sec & 11 hours \\ \hline
\end{tabular}
\caption{Condition of calculation related to 
the DE formula and the extrapolation, 
as well as the number of FPGA boards used in calculation and precision, 
and elapsed time.
$N$ is the number of grids per dimension, 
$h$ is a mesh size of the transformed variable in DE, and 
$\rho$ is a parameter variable used for the extrapolation.
\vspace{-0.3cm}
}
\label{tab:condition}
\end{center}
\end{table}

%
%
We computed a real part of $I$, $I_{re}$, in unphysical region,  
and both $I_{re}$ and an imaginary part, $I_{im}$, 
in physical region. 
Figure~\ref{fig:ssI} shows 
the behavior of our numerical results of $I$ as a function of $p^2$.
There are features which should be pointed out. 
First feature is that $I$ diverges at 
$p^2=1$
and $4$.
Around them,  
the numerical computation becomes very hard.
For example, the extrapolated results, 
$I_{re}(p^2=3.99)$ is not well
converged due to the severe singularity 
in the integration domain which causes the cancellation between positive and negative values.
Second feature is that there are points which a sign for $I$ changes at
$p^2 \sim 2$ and $2.5$ for $I_{re}$, and $p^2 \sim 3.5$ for $I_{im}$.
Last feature is that $I$ converges to zero for larger $p^2$.
These features are similar to those obtained 
in Ghinculov~\cite{Ghinculov_1996}. %
\footnote{The plot of $I_{re}$ and $I_{im}$ 
shown in Figure~5 of Ghinculov~\cite{Ghinculov_1996} 
seems to replace each other, 
because $I_{re}$ should not be zero if $p^2 < 1$. 
}

%
%
%
%


{
\setlength\abovecaptionskip{-5pt}
\begin{figure}[ht]
\begin{center}
\includegraphics[scale=0.35,angle=-90]{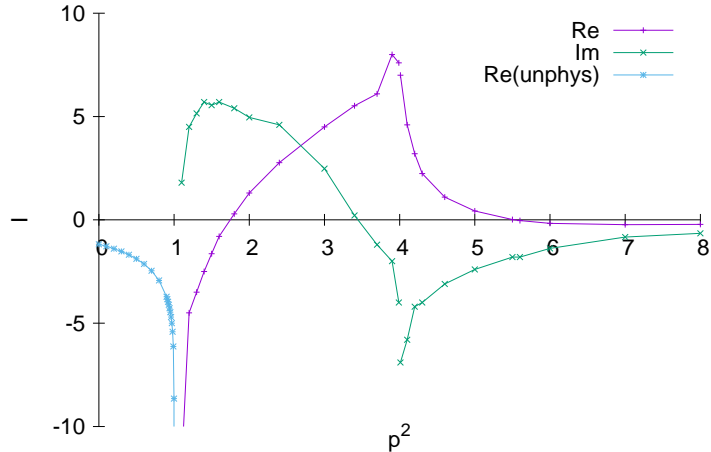}
\caption{Numerical results of $I$($I_{re}$ and $I_{im}$) 
as a function of $p^2$.
Purple and green denote $I_{re}$ and $I_{im}$ for physical region,
and blue denotes $I_{re}$ for unphysical region.
\vspace{-0.5cm}
}
\label{fig:ssI}
\end{center}
\end{figure}
}
Table~\ref{tab:condition} also shows the number of FPGA boards,  
precision used in calculation, and elapsed time per point $p$. 
By using our system with 64 FPGA boards, for physical region, 
a set of $I$ ($I_{re}, I_{im}$) with 24 different values of $\beta$ 
which is required for the extrapolation can be calculated in 11 hours 
which is acceptable. 
In order to increase the efficiency of calculation,
we performed integrations of 
$I_{re}$ and  $I_{im}$ with 8 different values of $\beta$ at the same time
(that is, 16 integrations at the same time).  
The reason why such calculation is possible is that  
PE has registers enough to store data,
and $C$ and $D$ are common so that they can be reused to
save computation. 
The calculation time of this method is about 4 times longer than 
that of a single integration case 
in which  $I_{re}$ or  $I_{im}$ is individually integrated with 
a single value of $\beta$.
Thus, in using this method, four times higher computational efficiency can be 
obtained.

\section{Summary and discussion}

We have been developing an accelerator system dedicated for
multi-precision arithmetic operators.
The current system is GRAPE9-MPX cluster system with 64 FPGA boards
on 8 host computers. 
Our dedicated processor, including PE with arithmetic logic units 
for MP4/MP6/MP8, is implemented on the FPGA boards.
We also developed a compiler system which enables us to accelerate
our application program on GRAPE9-MPX without any change in the program 
except for putting pragma just before double loops to be accelerated.
It should be noted that the current version of Goose can generate 
MPI API, and OpenCL API and kernels which has the DD format,
so that we can use a cluster system and GPGPU.
We performed an evaluation of a 3-loop self energy diagram with mass parameters 
used in Ghinculov~\cite{Ghinculov_1996}
and showed that our numerical method (DE-DCM + GRAPE9-MPX) can reproduce 
the behavior of $I$ seen in Ghinculov~\cite{Ghinculov_1996}.

Note that we need to discuss about error of numerical results 
in Fig.~\ref{fig:ssI}.
For unphysical region, the error is reasonable, 
but for physical region,  
the numerical error estimated from the extrapolation 
is 3-digit even for the best case and 1-digit for the worst case. 
The reason for the large error in physical region 
is probably due to the severe singularity appearing in the integral domain
which causes large cancellation.
One way to improve the accuracy is to tune parameters of the integration method,
for example, increasing the number of grids and/or reducing mesh size.
Another way is to divide the integral domain into sub domains 
and perform integration for each sub domain. 
In both cases, a large scale computation in multi-precision arithmetic 
such as quadruple/hexuple/octuple precision will play an important role.
The results will appear in future paper.

%
%
%

%
%
%


%
%
%
\section*{Acknowledgment}
We thank Prof. Kiyoshi Kato for fruitful discussion and comments.
This work is further supported by Grant-in-Aid for Scientific Research 
(15H03602, 15H03668, and 17K05428) of JSPS, 
and the Large Scale Simulation Program No. 16/17-21 of KEK.

\section*{References}

\bibliography{biblist}

\providecommand{\newblock}{}
\begin{thebibliography}{10}
\expandafter\ifx\csname url\endcsname\relax
  \def\url#1{{\tt #1}}\fi
\expandafter\ifx\csname urlprefix\endcsname\relax\def\urlprefix{URL }\fi
\providecommand{\eprint}[2][]{\url{#2}}

\bibitem{HAHN1999153}
{Hahn} T and {P\'erez-Victoria} M 1999 {\em Computer Physics Communications\/}
  {\bf 118} 153 -- 165

\bibitem{Wang_2004}
{Wang} J~X 2004 {\em Nucl.Instrum.Meth.\/} {\bf A534} 241--245

\bibitem{Belanger_2006}
{B\'elanger} G, {Boudjema} F, {Fujimoto} J, {Ishikawa} T, {Kaneko} T, {Kato} K
  and {Shimizu} Y 2006 {\em Phys. Rept.\/} {\bf 430} 117--209

\bibitem{Cullen_2011}
{Cullen} G, {Guillet} J~P, {Heinrich} G, 
  {Reiter} T and {Rodgers} M 2011 {\em Computer Physics Communications\/} {\bf
  182} 2276--2284

\bibitem{Hameren_2011}
{van Hameren} A 2011 {\em Computer Physics Communications\/} {\bf 182}
  2427--2438

\bibitem{Carrazza_2016}
{Carrazza} S, {Ellis} R~K and {Zanderighi} G 2016 {\em Computer Physics
  Communications\/} {\bf 209} 134--143

\bibitem{Donner_2017}
{Denner} A, {Dittmaier} S and {Hofer} L 2017 {\em Computer Physics
  Communications\/} {\bf 212} 220--238

\bibitem{Borowka_2017}
{Borowka} S, {Heinrich} G, {Jahn} S, {Jones} S~P, {Kerner} M and {Schlenk} J
  2017 {\em Journal of Physics: Conf. Series\/} {\bf 920}

\bibitem{Yuasa_2012}
{Yuasa} F, {de Doncker} E, {Hamaguchi} N, {Ishikawa} T, {Kato} K, {Kurihara} Y
  and {Fujimoto} J 2012 {\em Computer Physics Communications\/} {\bf 183}
  2136--2144

\bibitem{TM74}
{Takahasi} H and {Mori} M 1974 {\em Publications of the Research Institute for
  Mathematical Sciences\/} {\bf 9} 721–741

\bibitem{Doncker_2018}
{de Doncker} E, {Shimizu} Y, {Fujimoto} J and {Yuasa} F 2018 {\em Computer
  Physics Communications\/} {\bf 224} 164--185

\bibitem{GMP}
{GMP} \urlprefix\url{https://gmplib.org/}

\bibitem{QDLIB}
{High-Precision Software Directory}
  \urlprefix\url{http://crd-legacy.lbl.gov/~dhbailey/mpdist/}

\bibitem{MPFR}
{MPFR} \urlprefix\url{http://www.mpfr.org/}

\bibitem{Sugimoto_1990}
{Sugimoto} D, {Chikada} Y, {Makino} J, {Ito} T, {Ebisuzaki} T and {Umemura} M
  1990 {\em Nature\/} {\bf 345} 33--35

\bibitem{Makino_2006}
{Makino} J 2006 {\em Computing in Science \& Engineering\/} (8) 30--40

\bibitem{Daisaka_2011}
{Daisaka} H, {Nakasato} N, {Makino} J, {Yuasa} F and {Ishikawa} T 2011 {\em
  Procedia Computer Science\/} {\bf 4} 878--887

\bibitem{Nakasato_2012}
{Nakasato} N, {Daisaka} H, {Fukushige} T, {Kawai} A, {Makino} J, {Ishikawa} T
  and {Yuasa} F 2012 {\em Embedded Multicore Socs (MCSoC), 2012 IEEE 6th
  International Symposium on\/}  75--83

\bibitem{Motoki_2014}
{Motoki} S, {Daisaka} H, {Nakasato} N, {Ishikawa} T, 
  T, {Kawai} A and {Makino} J 2014 {\em Journal of Physics: Conference
  Series\/} {\bf 608}

\bibitem{Daisaka_2015}
{Daisaka} H, {Nakasato} N, {Ishikawa} T and {Yuasa} F 2015 {\em Procedia
  Computer Science\/} {\bf 51} 1323--1332

\bibitem{AlteraArriaVboard}
{Altera Co} {\it Arria V Device Handbook}
  \urlprefix\url{www.altera.co.jp/literature/hb/arria-v/arriav_handbook.pdf}

\bibitem{Nakasato_2009}
{Nakasato} N and {Makino} J 2009 {\em IEEE International Conference on Cluster
  Computing and Workshops\/} pp 1--9

\bibitem{Hida_2001}
{Hida} Y, {Li} X and {Bailey} D 2001 {\em {Proceedings of the 15th Symposium on
  Computer Arithmetic}\/} pp 155--162

\bibitem{Ghinculov_1996}
{Ghinculov} A 1996 {\em Physics Letters B\/} {\bf 385} 279--283

\bibitem{Wynn_1956}
{Wynn} P 1956 {\em Mathematical Tables and Aids to Computing\/} {\bf 10} 91--96

\end{thebibliography}

\end{document}